\documentclass[%
reprint,
superscriptaddress,
amsmath,
amssymb,
aps,
prl,
]{revtex4-2}

\usepackage{graphicx}
\usepackage{dcolumn}
\usepackage{bm}
\usepackage{bbold}
\usepackage[hypertexnames=false]{hyperref}
\hypersetup{
	colorlinks = true,
	allcolors = blue,
}

\newcommand{\ket}[1]{\left| #1 \right\rangle}

\newcommand{\ketbra}[2]{\left| #1 \right\rangle \left\langle #2 \right|}

\begin{document}
	\title{Mach-Zehnder interferometer for in-situ characterization of atom traps}
\author{A. Wolf}
\email{a.wolf@dlr.de}
\affiliation{German Aerospace Center (DLR), Institute of Quantum Technologies, 89081 Ulm, Germany}
\affiliation{Institut f\"ur Quantenphysik and Center for Integrated Quantum Science and Technology (IQST), Universit\"at Ulm, 89069 Ulm, Germany}
\author{M. A. Efremov}
\affiliation{German Aerospace Center (DLR), Institute of Quantum Technologies, 89081 Ulm, Germany}
\affiliation{Institut f\"ur Quantenphysik and Center for Integrated Quantum Science and Technology (IQST), Universit\"at Ulm, 89069 Ulm, Germany}
\date{\today}
\begin{abstract}
	Manipulating cold atoms in traps is a key tool for numerous realizations of quantum simulators and quantum sensors. They require accurate modeling and characterization of the underlying trapping potentials. We introduce a technique based on the Mach-Zehnder interferometer for in-situ characterization of weakly anharmonic potentials. By simulating the interferometer in an optical dipole trap, we can accurately determine its trap frequency and upper bounds onto anharmonicity magnitudes.
\end{abstract}
\maketitle
\noindent\textit{Introduction.--}
The possibility to control the motion of ultracold atoms with magnetic or laser-generated potentials has given rise to the field of atomtronics~\cite{Amico.2021} where it is used to realize matter-wave circuits, {\it i.e.} networks for atomic currents. Closely related to this is the concept of guided atom interferometers where atoms are held in potentials with beam splitters and mirrors realized by inserting junctions into guides~\cite{Dumke.2002}, deforming the potentials~\cite{Petrucciani.2025}, or diffraction by a light wave~\cite{Xu.2019,Panda.2024,Gebbe.2021,Horikoshi.2007,Burke.2008,Sapiro.2009,McDonald.2013b,McDonald.2013,Masi.2021,Moan.2020,Krzyzanowska.2023,Beydler.2024}. Trapping atoms either for some part of~\cite{Xu.2019,Panda.2024,Gebbe.2021} or for the entire interferometric sequence~\cite{Horikoshi.2007,Burke.2008,Sapiro.2009,McDonald.2013b,McDonald.2013,Moan.2020,Masi.2021,Krzyzanowska.2023,Balland.2024,Beydler.2024,Petrucciani.2025} is used to increase the momentum transfer~\cite{McDonald.2013,Gebbe.2021}, enclosed area~\cite{Moan.2020,Gebbe.2021,Krzyzanowska.2023,Beydler.2024}, or interrogation time~\cite{Xu.2019,Panda.2024,Petrucciani.2025}. These make the interferometer more sensitive to external fields, forces, and rotation, as well as highly miniaturized~\cite{Beydler.2024}. Hence, control and accurate characterization of the guiding potential are key issues~\cite{Amico.2021} to improve interferometer performance. In this Letter, we propose an in-trap interferometric scheme to accurately quantify a weakly anharmonic trap. Here we do not require any mechanism of momentum transfer, adiabatic trap deformation, or switching off the trap. The trap frequency and upper bounds for the magnitudes of anharmonicities (cubic and quartic) are determined from the interferometer signal.

Atoms can be confined with optical dipole~\cite{Grimm.2000} or magnetic traps~\cite{Foot.2011}, including atom chips~\cite{Folman.2002,LewoczkoAdamczyk.2009,Keil.2016}. A common way to measure the trap frequencies and possibly anharmonicity amplitudes relies on the ballistic technique where a classical oscillatory trajectory of atoms inside the trap is excited by displacing the trap~\cite{Beydler.2024}, loading the trap off its center~\cite{Horne.2017,Pfeiffer.2023}, or transferring momentum to the atoms~\cite{Horikoshi.2007,Masi.2021,Beydler.2024}. Alternative techniques are based on parametric heating where the response of the trapped gas to periodic perturbation is measured~\cite{Jauregui.2001,Moon.2010,Lauber.2011,Makhalov.2015}.

\begin{figure}[t]
	\centering
	\includegraphics[width=245.99557pt,height=248.07623pt]{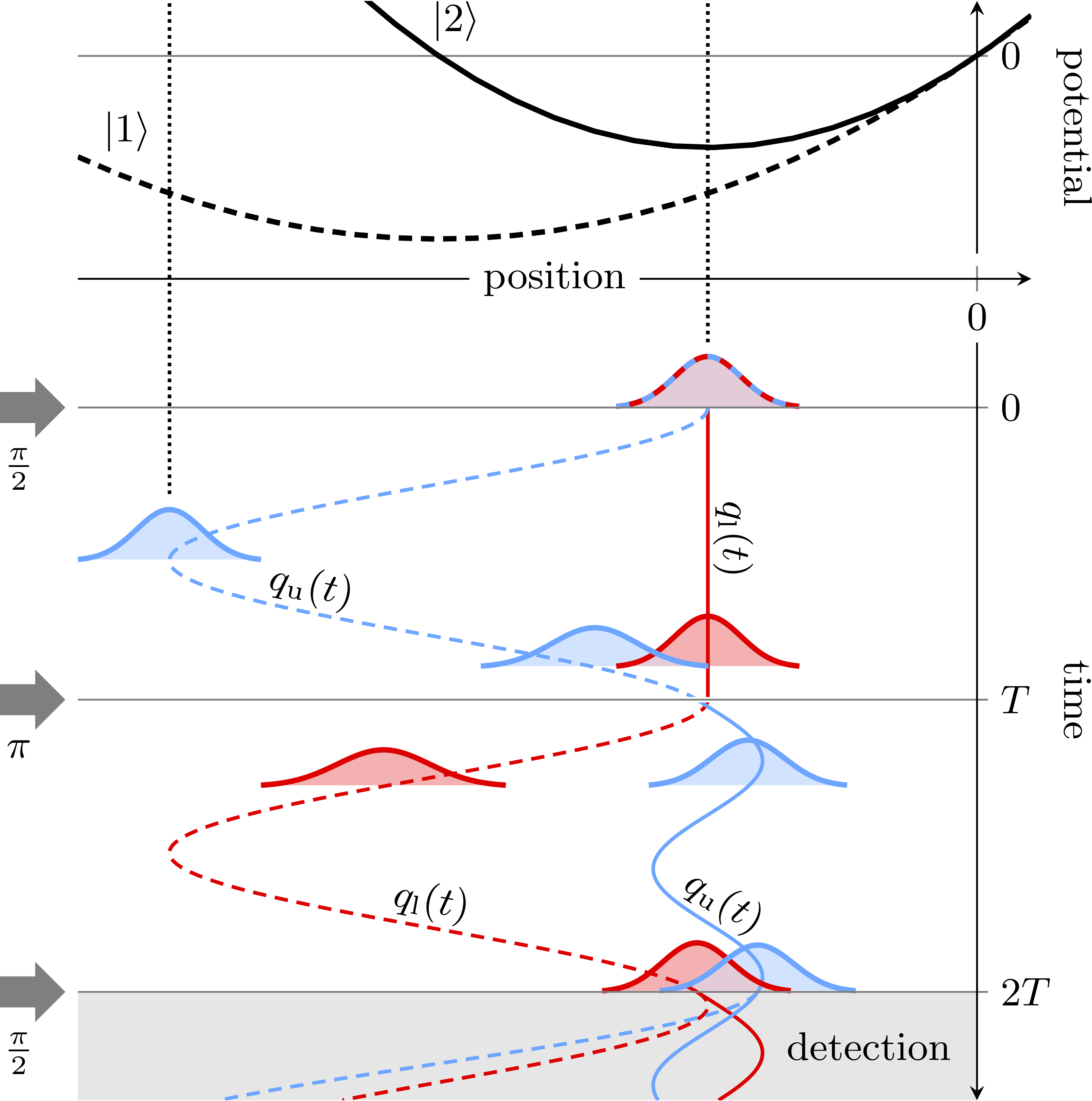}
	\caption{\label{fig:scheme1D}
		(top) Harmonic trapping potentials of two internal states $\ket{1}$ (dashed) and $\ket{2}$ (solid) with spatial shift between the trap centers (minima). (bottom) Interferometer pulses (thick arrows) separated by time $T$ transfer atoms between the two states, automatically leading to moving and, generally, breathing wavepackets. The shown Mach-Zehnder configuration leads to a periodic two-path-interference of an upper (blue) and lower (red) wavepacket whenever the positions $q$ of their center-of-mass are sufficiently close.
	}
\end{figure}

\noindent\textit{Scheme in one dimension.--}
We present a method to measure the trap frequencies by means of the atomic Mach-Zehnder interferometer (MZI) depicted in Fig.~\ref{fig:scheme1D}.
The scheme requires an atom with two trapped internal states $\ket{\alpha}$, $\alpha = 1,2$ that have different minima of the trapping potentials. This is the case for the state-dependent potentials $V_\alpha(z) = m \omega_\alpha^2 z^2 / 2 - F_\alpha z$, consisting of a harmonic trap and a constant force, where either the two frequencies $\omega_\alpha$ or the two forces $F_\alpha$ have to differ, for the trap minima at $z_{0,\alpha} = F_\alpha / (m \omega_\alpha^2)$ to be separated. Examples are either two trapped states in a magnetic trap ($\omega_1 \neq \omega_2$) with gravitational sag ($F_1 = F_2$), or two differently levitated states ($F_1 \neq F_2$) in an optical dipole trap ($\omega_1 \approx \omega_2$).

An atomic cloud is initially prepared at rest in the state $\ket{2}$ and thus sees the potential $V_2(z)$, shown by the solid line in Fig.~\ref{fig:scheme1D}. At $t = 0$, a short light pulse populates the state $\ket{1}$ and creates a new wavepacket that starts to move in the potential $V_1(z)$, shown by dashed line in Fig.~\ref{fig:scheme1D}. Since $V_1(z)$ and $V_2(z)$ are confining potentials, the two wavepackets periodically overlap and can be brought to interfere with each other by again transferring population between the states. The trap frequencies are extracted from the measured populations of the internal states after the last pulse, $t = 2T$, Fig.~\ref{fig:scheme1D}.

In the MZI each wavepacket spends the same amount of time $T$ in both states due to the central $\pi$-pulse, at $t = T$, that inverts the population of the states. Consequently, the MZI is insensitive to the ground-state energy of each harmonic potential or any static background field. Moreover, due to the periodic motion alone, no momentum kicks are required to achieve spatial separation and recombination of the wavepackets. Hence, population transfer between the internal states can be realized with radio-frequency fields or with co-propagating beams in the microwave regime via Raman transitions, leading to negligible momentum transfer during pulses.

To model the one-dimensional (1D) MZI, we approximate the pulses as instantaneous perfect transfers of the population and assume the initial state to be the Gaussian ground state of the potential $V_2(z)$. The Gaussian wavepackets evolve according to the Newton and Ermakov equations for their center-of-mass position $q(t)$ and width $\sigma(t)$, respectively~\cite{Haas.2013}. The population of state $\ket{2}$ after the last pulse then reads
\begin{equation}\label{eq:observable}
	P_2(T) = \frac{1}{2} \Big\{1 + C(T) \cos\left[\varphi(T) + \Delta\Phi_{\text{P}}\right]\Big\} \,.
\end{equation}
The contrast $C(T)$ and phase $\varphi(T)$ are defined via the overlap $C(T) \mathrm{e}^{\mathrm{i}\varphi(T)} = \int \mathrm{d}z \psi_\text{u}^\ast(z,2T) \psi_\text{l}(z,2T)$ of the two wavepackets at the last pulse, $t = 2T$. The phase $\Delta\Phi_{\text{P}}$ is the difference of two phases imprinted onto the two arms of the MZI during the light pulses. More details on the analytic results are summarized in the End Matter~\footnote{See Supplemental Material at [URL will be inserted by publisher] for more details on the analytic derivation of the interferometer signal and the extraction of trap frequencies from the signal of the MZI inside an optical dipole trap.}.
\begin{figure}[t]
	\centering
	\includegraphics[width=246.00125pt,height=240.34258pt]{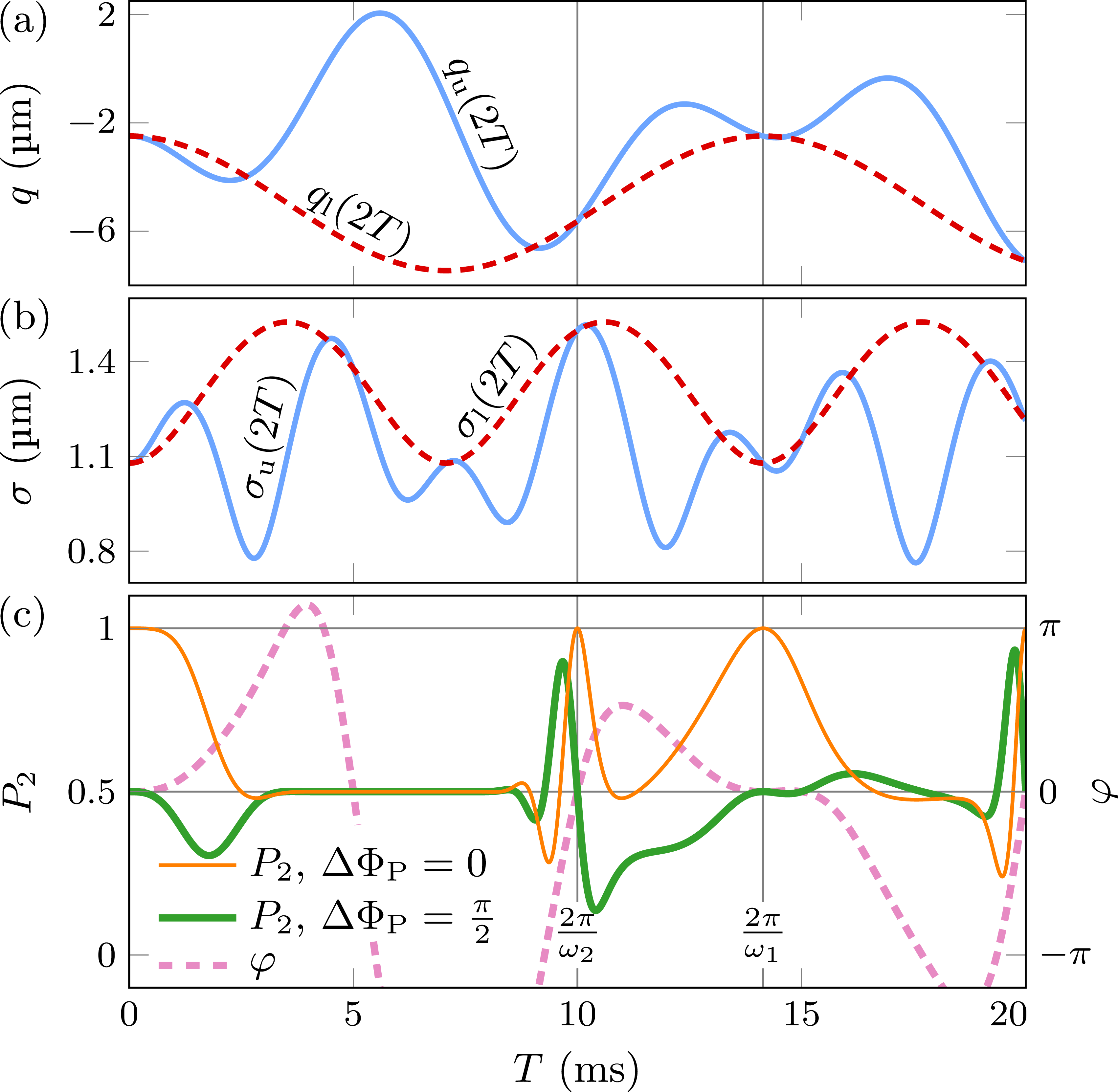}
	\caption{\label{fig:signal1D}
		(a) Positions $q$ and (b) widths $\sigma$ of the two wavepackets at the time of the final pulse $2T$ as a function of the time $T$ between the pulses. Here the parameters listed in Tab.~\ref{tab:example} are used. (c) Signal $P_2$ of the interferometer for two pulse phase contributions $\Delta \Phi_\text{P}$. When $T$ matches the trap periods $2 \pi / \omega_2$ or $2 \pi / \omega_1$ (vertical lines) the wavepackets are identical at the last pulse and the phase $\varphi$ vanishes.
	}
\end{figure}
\begin{table}[b]
	\caption{\label{tab:example}
		Parameters of our study case. Here $m$ denotes the mass of $^{87}$Rb atom with the typical trapped internal states $\ket{1} = \ket{F = 2,m_\text{F} = 1}$ and $\ket{2} = \ket{F = 2,m_\text{F} = 2}$. The applied acceleration $g \equiv F / m$ is in units of the Earth acceleration $g_\text{E} = 9.81~\mathrm{m \cdot s^{-2}}$.
	}
	\begin{ruledtabular}
		\begin{tabular}{c|ccccc}
			axis & $\omega_2 / (2 \pi)$~(Hz) & $\omega_1 / (2 \pi)$~(Hz) & $F_1$ & $F_2$ & $g / g_\text{E}$\\\colrule
			$z$ & \rule{0pt}{3ex} $100$ & $100 / \sqrt{2}$ & $-mg$ & $-mg$ & $0.1$\\\colrule
			$x$ & \rule{0pt}{3ex} $177.77$ & $177.77 / \sqrt{2}$ & 0 & 0 & -\\
			$y$ & \rule{0pt}{3ex} $277.77$ & $277.77 / \sqrt{2}$ & 0 & 0 & -
		\end{tabular}
	\end{ruledtabular}
\end{table}

When the positions of the wavepackets at the last pulse, $t = 2T$, Fig.~\ref{fig:signal1D}(a), are farther apart than the respective widths, Fig.~\ref{fig:signal1D}(b), there is no significant overlap and the contrast drops to zero. In this case the interferometer signal, Eq.~\eqref{eq:observable}, flattens at $P_2(T) = 0.5$, as seen in Fig.~\ref{fig:signal1D}(c) for $4~\text{ms} < T < 8~\text{ms}$. However, the signal (thin orange line) displays a peak when $T$ is close to an integer number of the trap periods $\mathcal{T}_1 = 2 \pi / \omega_1$ and $\mathcal{T}_2 = 2 \pi / \omega_2$ (vertical lines). At these times both the positions $q_l(2\mathcal{T}_\alpha)$, $q_u(2\mathcal{T}_\alpha)$ and widths $\sigma_l(2\mathcal{T}_\alpha)$, $\sigma_u(2\mathcal{T}_\alpha)$ of the wavepackets coincide, {\it i.e.} the interferometer perfectly closes~\footnote{Technically, a closed interferometer requires that the wavepackets have perfect overlap in both position and momentum space, which is satisfied in our case when $q_\text{u}(2\mathcal{T}_\alpha) = q_\text{l}(2\mathcal{T}_\alpha)$ and $\sigma_\text{u}(2\mathcal{T}_\alpha) = \sigma_\text{l}(2\mathcal{T}_\alpha)$, simultaneously.}. The widths of the peaks scale quadratically in the distance $|z_{0,2} - z_{0,1}|$ and become narrower for larger separations because the wavepackets spend more time of a period apart.

The two types of peaks at $\mathcal{T}_1$ and $\mathcal{T}_2$ originate from two different trajectories. Figure~\ref{fig:scheme1D} depicts the case $T = 13.5~\text{ms}$ which is slightly smaller than $\mathcal{T}_1 \approx 14.14~\text{ms}$. For $T = \mathcal{T}_1$, the $\pi$-pulse is applied when the motion of the upper arm (blue) is at its turning point in the trap $V_1(z)$. The atoms are then transferred back to rest in the trap $V_2(z)$, while the lower arm (red) performs the identical motion in $V_1(z)$. Note that if $T = \mathcal{T}_1$, the pulses are always applied when the atoms are back at their initial position. The case of $T = \mathcal{T}_2$ is very different, as the atoms of the upper arm are transferred back to $V_2(z)$, while they are somewhere in motion. However, the interferometer still closes, because the atoms complete one full oscillation in $V_2(z)$ during the second half of the interferometer, while the atoms in the lower arm propagate to the same point.

To determine the period $\mathcal{T}_2$ of the trap in which the atoms are initially prepared, one (i) sets $\Delta \Phi_\text{P} = \pi / 2$ and (ii) measures the times $T$ at which the signal crosses the value $P_2(T) = 0.5$ with a large slope. This occurs periodically at $T = n \mathcal{T}_2$ with integer $n$, as shown in Fig.~\ref{fig:signal1D}(c) for $n = 1$ (thick green line). To estimate the sensitivity of this approach we expand contrast and phase around $n \mathcal{T}_2$, $\varphi(T) \approx s (T - n \mathcal{T}_2)$ and $C(T) \approx 1 + \mathcal{O}\left[(T - n \mathcal{T}_2)^2\right]$, where $s$ is dominantly given by the center-of-mass motion
\begin{equation}\label{eq:slope}
	s_\text{CM} = 4 \frac{E}{\hbar} \sin^4\left(\frac{\pi n}{r}\right) \left[r^2 + \cot^2\left(\frac{\pi n}{r}\right) \right] \,.
\end{equation}
Here $r = \omega_2 / \omega_1$ and $E = m \omega_1^2 (z_{0,2} - z_{0,1})^2 / 2$ is the energy given to the center-of-mass motion of the upper wavepacket in the first interval. Applying Gaussian error propagation relates the uncertainty $\Delta \omega_2$ of $\omega_2$ to the one $\Delta P_2(T)$ of $P_2(T)$, $\Delta \omega_2 = \Delta P_2(T) / |\partial P_2(T) / \partial \omega_2|$. The latter reads $\Delta P_2(T) = \sqrt{P_2(T) [1 - P_2(T)] / N}$, where $N$ is the number of measurements. Inserting the expansions for phase and contrast yields the relative uncertainty $\Delta \omega_2 / \omega_2 = 1 / \left(s n \mathcal{T}_2 \sqrt{N}\right)$ when the interferometer is operated at $T = n \mathcal{T}_2$ with $\Delta \Phi_\text{P} = \pi / 2$. The sensitivity of our method increases for larger slopes $s$, that is for larger $\sin^4(\pi n / r)$, Eq.~\eqref{eq:slope}. For the parameters listed in Tab.~\ref{tab:example} the slope $s$ is large at $n = 5$ ($T= 50~\text{ms}$) and one obtains $\Delta \omega_2 / \omega_2 \approx 4.3 \times 10^{-6}$ with $5000$ atoms and $100$ shots, {\it i.e.} $N = 5 \times 10^5$. If coherence time permits, more orders of magnitude can be gained by increasing $n$ provided large $\sin^4(\pi n / r)$. Note, this method only works for different trap frequencies, since $s_\text{CM} = 0$ for $r = 1$.

\noindent\textit{Extension to three dimensions.--}
We generalize our scheme to a three-dimensional (3D) scenario where the trapping potentials are separated by a non-zero distance along one axis. This is achieved by aligning the force with one of the trap's principal axes, resulting in the potential $V_\alpha(\mathbf{x}) = \sum_{i = 1}^3 m \omega_{\alpha,x_i}^2 x_i^2 / 2 - F_\alpha z$ with $\mathbf{x} = (x_1,x_2,x_3) = (x,y,z)$~\footnote{If the force is not aligned with a principal axis, all three center-of-mass motions are excited when the atoms are transferred to the other trap. In the general case where the trap frequencies of the three one dimensional traps have no common multiple, the atoms then orbit around their starting position and never come back.}. The contrast $C(T) = C_x(T) C_y(T) C_z(T)$ and the phase $\varphi(T) = \varphi_x(T) + \varphi_y(T) + \varphi_z(T)$ are then given by the contrasts $C_{x_i}(T)$ and the phases $\varphi_{x_i}(T)$ describing the one-dimensional contribution of the $x_i$-direction. 

\begin{figure}[t]
	\centering
	\includegraphics[width=246.00142pt,height=118.52002pt]{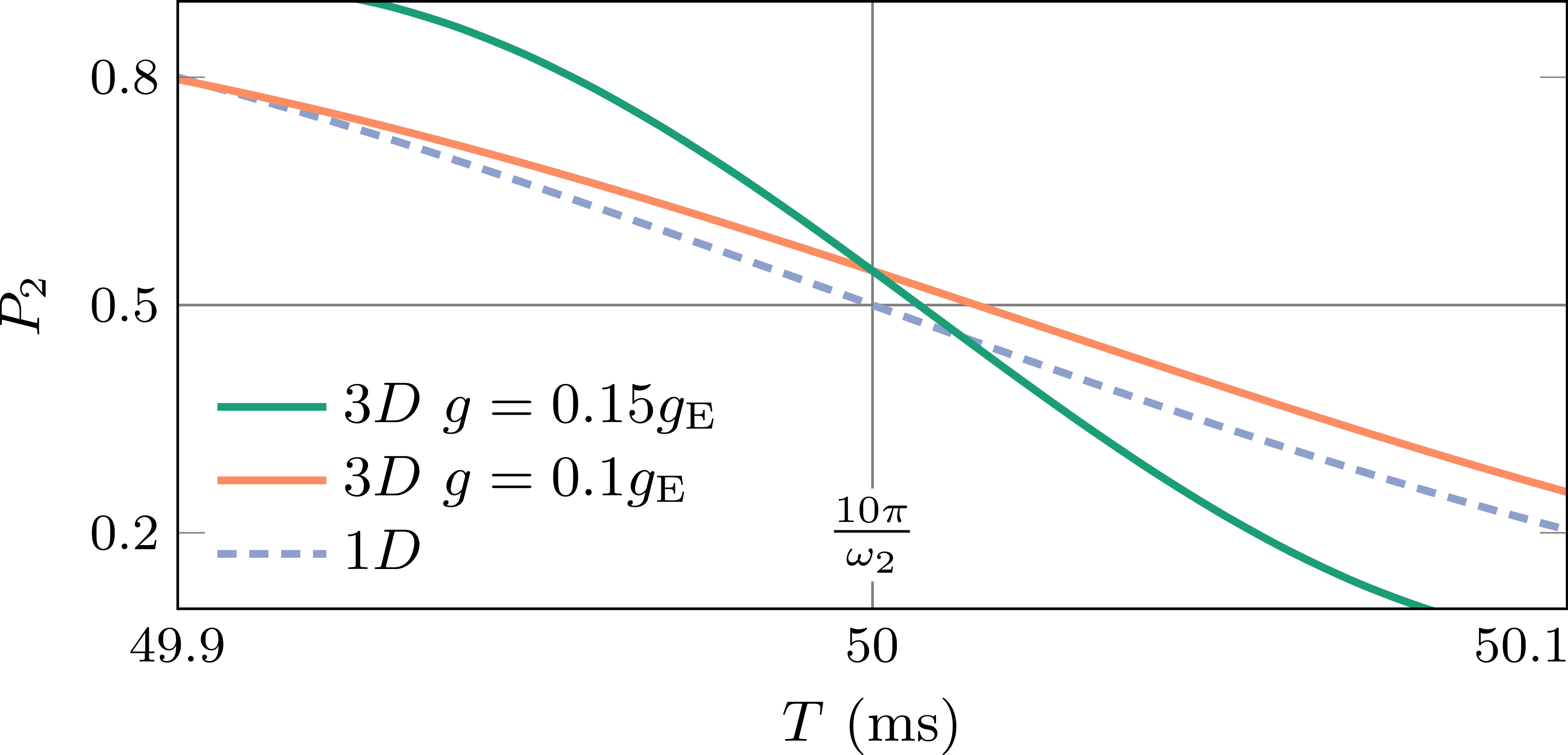}
	\caption{\label{fig:signal3D}
		Signal $P_2$ of the 1D and 3D MZI for the time $T$ close to $10 \pi / \omega_2$. The parameters listed in Tab.~\ref{tab:example} and the phase difference $\Delta \Phi_\text{P} = \pi / 2$ are used.
	}
\end{figure}

Our 1D scheme to measure the period $\mathcal{T}_2 = 2 \pi / \omega_{2,z}$ in the longitudinal direction $z$ becomes problematic, when the breathing motion of the atomic cloud in the transversal directions $x$ or $y$ has a different periodicity compared to the $z$ direction. For example, if both $\omega_{1,x} \neq \omega_{2,x}$ and $\omega_{1,x} \neq \omega_{1,z}$, the revivals of the contrast are not perfect anymore, because even when the interferometer is closed  along the $z$-direction, it is not closed in the transversal directions, since at $T=n\mathcal{T}_2$, $C_z(n\mathcal{T}_2) = 1$, but  $C_{x,y}(n\mathcal{T}_2)<1$ and $\varphi_{x,y}(n \mathcal{T}_2) \neq 0$. In Fig.~\ref{fig:signal3D}, the signal $P_2(T)$ of the 3D MZI (solid lines) is contrasted to the one of the 1D MZI (dashed line). For $T = 5 \mathcal{T}_2$, $P_2(T) \neq 0.5$ due to the nonzero phases coming from the transversal breathing motion. This is not easily accounted for since the phases depend on unknown transversal trap frequencies.

A solution to this problem is repeating the experiment with two different magnitudes of the applied forces, corresponding to different separations $|z_{0,2} - z_{0,1}|$ between the longitudinal traps, and determine the time $T$ when the two signals cross, as shown in Fig.~\ref{fig:signal3D}. This should occur at $T = n \mathcal{T}_2$, because $C_{x,y}(T)$ and $\varphi_{x,y}(T)$ are independent of the applied forces, while the slope $s_\text{CM}$, Eq.\eqref{eq:slope}, with which $\varphi_z(T)$ approaches zero at $T = n \mathcal{T}_2$, scales as $(z_{0,2} - z_{0,1})^2$. 

\noindent\textit{Pulse effects.--}
Two important aspects were neglected earlier by using perfect instantaneous pulses. First, the nonzero times $\tau_{\pi}$ and $\tau_{\pi / 2}$ of the $\pi$- and $\pi / 2$-pulses result in a total propagation time $2T + 2\tau_{\pi / 2} + \tau_{\pi}$. This leads to a small systematic horizontal shift of the whole signal. With the method presented above, it is sufficient to determine two neighboring crossings of the value $P_2(T) = 0.5$. The distance between them then defines the period $\mathcal{T}_2$. Second, because of the potentials $V_1(\mathbf{x})$ and $V_2(\mathbf{x})$, the resonance condition for driving transitions between the two internal states depends on the position of the atom. According to Eq.~\eqref{eq:slope}, a larger distance $|z_{0,2} - z_{0,1}|$ raises the slope $s_\text{CM}$ and hence the precision of our method. However, this also increases the position-dependent detuning and makes the $\pi$- and $\pi / 2$-pulses less efficient.

To investigate the performance of the proposed MZI with respect to the applied forces, we simulate the 3D MZI, using a Rabi-coupled two-level Hamiltonian and box-shaped pulses without pulse phase ($\Delta \Phi_\text{P} = 0$), as detailed in the End Matter. The pulses are driven with the frequency such that the transition at the initial position [right vertical dotted line Fig.~\ref{fig:scheme1D}(top)] is resonant. After the first $\pi / 2$-pulse, the atoms in $\ket{1}$ start to move and the largest detuning they acquire is at the position of the other turning point [left vertical dotted line Fig.~\ref{fig:scheme1D}(top)]. Subtracting the level splitting $V_2(0,0,z_{0,2}) - V_1(0,0,z_{0,2})$ from $V_2(0,0,2 z_{0,1} - z_{0,2}) - V_1(0,0,2 z_{0,1} - z_{0,2})$ yields the detuning $\delta = 2 (F_2 \omega_1^2 - F_1 \omega_2^2)^2 / \left(\hbar m \omega_1^4 \omega_2^2\right)$.

\begin{figure}[t]
	\centering
	\includegraphics[width=246.00003pt,height=240.1652pt]{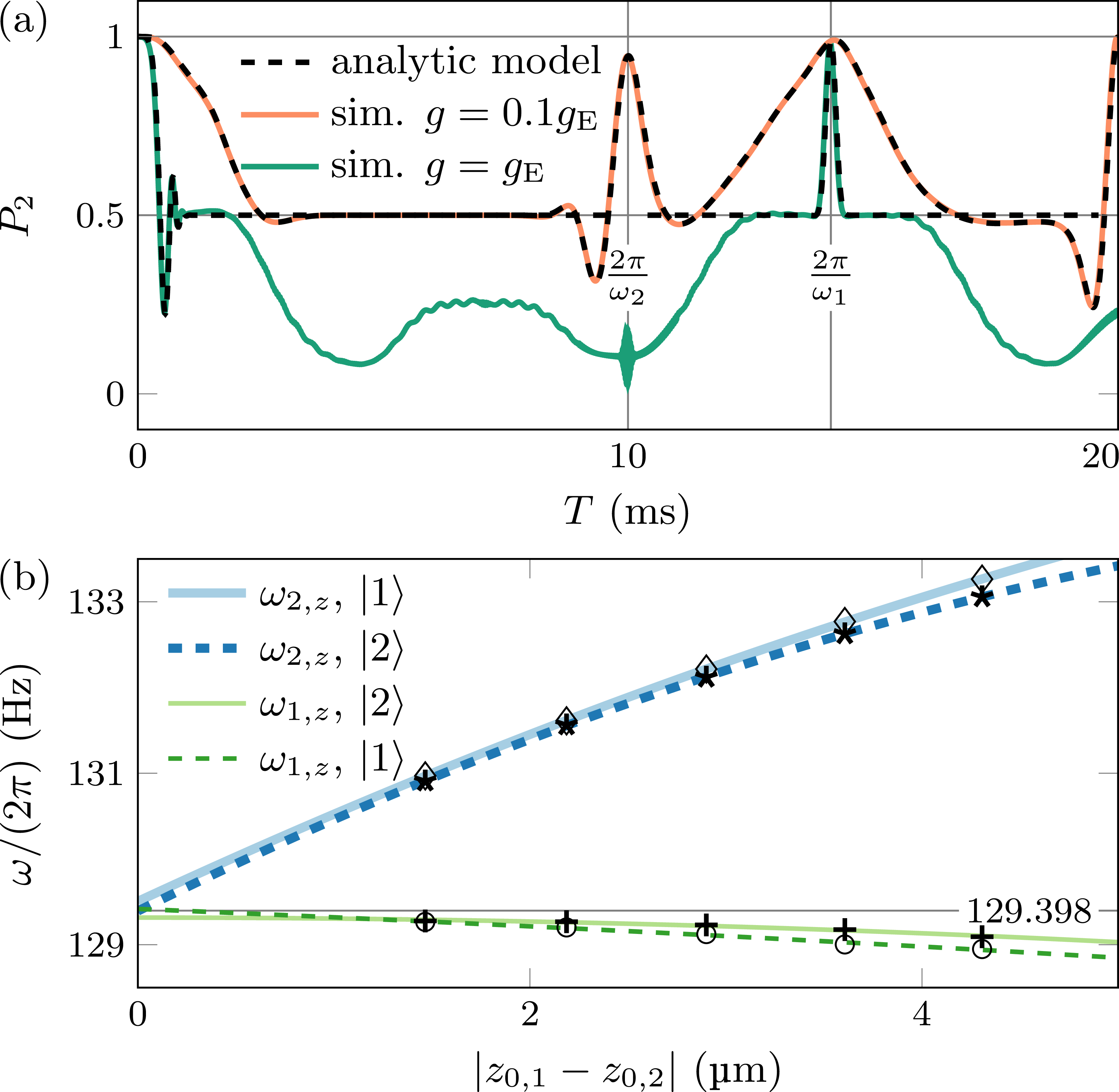}
	\caption{\label{fig:signalReal}
		(a) Interferometer signals $P_2$ as a function of the pulse separation $T$. The trap parameters are listed in Tab.~\ref{tab:example}. Solid lines are simulations of a Rabi-coupled two-level Hamiltonian including the influence of position-dependent detunings. (b) Trap frequencies $\omega$ and corresponding fits determined by simulating the 3D MZI inside an optical dipole trap for different trap separations $|z_{0,1} - z_{0,2}|$ and initial states.
	}
\end{figure}

For the Rabi frequency $\Omega = 2 \pi \times 25~\text{kHz}$, corresponding to $\tau_{\pi / 2} = 10~\text{\textmu s}$ and $\tau_{\pi} = 20~\text{\textmu s}$, and the parameters listed in Tab.~\ref{tab:example}, we obtain the small detuning $\delta \approx 0.042\,\Omega$ and Fig.~\ref{fig:signalReal}(a) shows that the 3D analytic model accurately describes our MZI. However, for a applied force one order of magnitude larger, {\it i.e.} $g = g_\text{E}$, the detuning becomes too large, $\delta \approx 4.25\,\Omega$, and the $\pi$-pulse fails to exchange the populations between the states $\ket{1}$ and $\ket{2}$. At $T = \mathcal{T}_2$, there are quick suppressed oscillations, Fig.~\ref{fig:signalReal}(a), due to the large slope, Eq.~\eqref{eq:slope}, and greatly reduced contrast. Conversely, at $T = \mathcal{T}_1$, the atoms come back to the initial position, Fig.~\ref{fig:scheme1D}, for all pulses and our analytic model remains very accurate. Hence, for large detunings, an alternative way for extracting the trap frequency is to identify the narrow quasi-Gaussian peaks at $n \mathcal{T}_1$ and determine the time interval between such peaks.

\noindent\textit{Application to optical dipole traps.--}
We now examine the 3D MZI with perfect pulses inside a crossed optical dipole trap. As considered in the End Matter, the trapping potential is modeled as two Gaussian laser beams propagating along the $x$- and $y$-axis with wavelength $1064~\text{nm}$, power $1~\text{W}$, and beam waist $100~\text{\textmu m}$~\cite{Grimm.2000}. The ${}^{87}$Rb atoms in the trap are pulled downwards by gravity with $F_1 = - m g_\text{E}$ such that the trap center is at $z_{0,1} \approx -13.72~\text{\textmu m}$ below the focus of the laser beams. Expanding the potential into a Taylor series at $z_{0,1}$ up to the second order results in the trap frequencies $\omega_{1,x} = \omega_{1,y} \approx 2 \pi \times 95.15~\text{Hz}$ and $\omega_{1,z} \approx 2 \pi \times 129.398~\text{Hz}$. Our goal is to determine $\omega_{1,z}$ with the 3D MZI and investigate the influence of small anharmonicities~\cite{Leonard.2012,LaRow.2025,Steiner.2025}. As $\ket{1}$ we choose the magnetically insensitive state $\ket{m_F = 0}$, while as $\ket{2}$ we choose a magnetically sensitive state $\ket{m_F \neq 0}$. In this way, a magnetic field gradient provides the force $F_2 = - m g_2$ with a controllable effective acceleration $g_2 \neq g_\text{E}$.

Due to the shape of the optical dipole potential, the trap frequencies experienced by the atoms in state $\ket{2}$ slightly depend on $g_2$. For example, at $g_2 = 0.7 g_\text{E}$ we obtain $z_{0,2} \approx -9.42~\text{\textmu m}$, $\omega_{2,x} = \omega_{2,y}\approx 2 \pi \times 96.11~\text{Hz}$ and $\omega_{2,z} \approx 2 \pi \times 133.48~\text{Hz}$. The difference of the transversal frequencies $\omega_{1,x} - \omega_{2,x}$ is too small for the corresponding breathing motion to be important. However, the corresponding difference in the longitudinal ones is large enough for the peaks in the signal $P_2(T)$ to be clearly distinguishable at $90~\text{ms} < T < 155~\text{ms}$. Since the interferometer can be started either in $\ket{1}$ or in $\ket{2}$, both trap frequencies $\omega_{1,z}$ and $\omega_{2,z}$ can be independently determined by performing the 3D MZI for $g_2 / g_\text{E} = 0.7, 0.75, 0.8, 0.85, 0.9$, as displayed in Fig.~\ref{fig:signalReal}(b)~\cite{Note1}. Starting the 3D MZI in the states $\ket{2}$ and $\ket{1}$ give us slightly different estimates for $\omega_{1,z}$, circles and crosses, and $\omega_{2,z}$, stars and diamonds. Moreover, both frequencies clearly display a dependence on the distance between the trap centers $|z_{0,1} - z_{0,2}|$, that is on $g_2$. For $\omega_{1,z}$ this entirely originates from small anharmonicities of $V_1(\mathbf{x})$. The frequency $\omega_{2,z}$ has a stronger dependence on the distance because the trap minimum $z_{0,2}$ changes with $g_2$. As outlined in the End Matter, we use the standard formula for the classical oscillation frequency in the presence of weak anharmonicities to justify the fitting functions (quadratic polynomials) shown by the solid lines in Fig.~\ref{fig:signalReal}(b). Extrapolating these lines to $|z_{0,1} - z_{0,2}| = 0$, where the effect of anharmonicities vanishes and the traps are identical, yields four estimations for $\omega_{1,z}$. These four values give the average $\omega_{1,z}^\text{av} \approx 2 \pi \times 129.408~\mathrm{Hz}$ with the relative deviation $1 - \omega_{1,z} / \omega_{1,z}^\text{av} \approx 7.7 \times 10^{-5}$ to the true value $\omega_{1,z}=2 \pi \times 129.398~\text{Hz}$. Finally, using the fit of $\omega_{1,z}$ given by the lower (green) solid line, we obtain the upper bounds $|\alpha| < 4.63 \times 10^{-16}~\mathrm{J \cdot m^{-3}}$ and $|\beta| < 5.63 \times 10^{-12}~\mathrm{J \cdot m^{-4}}$ for the magnitudes of the cubic $\alpha$ and quartic  $\beta$ anharmonic terms $\alpha (z - z_{0,1})^3$ and $\beta (z - z_{0,1})^4$ in $V_1(\mathbf{x})$. They are consistent with the values $\alpha \approx 2.76 \times 10^{-16}~\mathrm{J \cdot m^{-3}}$ and $\beta \approx -4.39 \times 10^{-12}~\mathrm{J \cdot m^{-4}}$ derived with the Taylor expansion.

\noindent\textit{Conclusion.--}
We have proposed a Mach-Zehnder atom interferometer between two trapped states to determine trap frequencies and shown that it offers an estimated relative uncertainty $\Delta \omega / \omega \approx 4.3 \times 10^{-6}$ for the trapping frequency $\omega = 2 \pi \times 100~\text{Hz}$ and interrogation time $2T = 100~\text{ms}$. Ballistic measurements have demonstrated a relative uncertainty $\Delta \omega / \omega \approx 7.8 \times 10^{-5}$ for the frequency $\omega \approx 2 \pi \times 3.8673~\text{Hz}$ and observation times up to $100~\text{s}$~\cite{Beydler.2024}. Such high accuracy improves the performance of atomic lenses~\cite{Ammann.1997,Pandey.2021} and precision atom interferometry~\cite{Tino.2014}, and enables the determination of thermodynamic properties of a trapped BEC~\cite{Krstajic.2025}, or estimations of trap parameters, {\it e.g.} the beam waist at the atomic position.

The authors of Ref.~\cite{Beydler.2024} determined the magnitudes of anharmonic terms, while we have currently obtained only upper bounds. To determine their values with our scheme, the analytic description needs to be extended by including anharmonic terms in the trapping potentials. The interferometer signal could then be fitted to an improved expression. Furthermore, the decoherence due to the interplay of temperature and anharmonicities~\cite{Afek.2017} and the effect of interatomic interactions~\cite{Kafle.2011,Fogarty.2013} can be studied.
\begin{acknowledgments}
	The authors gratefully acknowledge W. P. Schleich for fruitful discussions as well as the scientific support and HPC resources provided by the German Aerospace Center (DLR). The HPC system CARA is partially funded by the Saxon State Ministry for Economic Affairs, Labour and Transport and the Federal Ministry for Economic Affairs and Climate Action. The authors acknowledge support by the state of Baden-Württemberg through bwHPC and the German Research Foundation (DFG) through Grant No. INST 40/575-1 FUGG (JUSTUS 2 cluster). This work was supported by the Science Sphere Quantum Science of Ulm University and the Center for Integrated Quantum Science and Technology (IQST) for financial support. The research of the IQST is financially supported by the Ministry of Science, Research and Arts Baden-Württemberg.
\end{acknowledgments}
\providecommand{\noopsort}[1]{}\providecommand{\singleletter}[1]{#1}%
\clearpage
\section{End Matter}
\noindent\textit{Analytic model.--}
A driven two-level atom of mass $m$ in an external state-dependent potential $V_\alpha(\mathbf{x})$, with $\alpha=1,2$, is described by the Hamiltonian
\begin{equation}\label{eq:Hamiltonian}
	\begin{aligned}
		H =& -\frac{\hbar^2}{2 m} \nabla^2 \mathbb{1} + V_1(\mathbf{x}) \ketbra{1}{1}+V_2(\mathbf{x}) \ketbra{2}{2}\\
		+& \hbar \Omega(t) \cos\left[\omega_\text{d} t + \phi(t)\right] \sigma_x + \frac{\hbar \omega_\text{b}}{2} \sigma_z
	\end{aligned}
\end{equation}
with the Rabi frequency $\Omega(t)$ and the driving frequency $\omega_\text{d}$. Here $\phi(t)$ denotes the pulse phase and $\hbar \omega_\text{b}$ is the transition frequency between the internal states $\ket{1}$ and $\ket{2}$. Moreover, $\mathbb{1}$ denotes the identity $2\times 2$ matrix, whereas $\sigma_x = \ketbra{1}{2} + \ketbra{2}{1}$ and $\sigma_z = \ketbra{2}{2} - \ketbra{1}{1}$ are the Pauli matrices. Note that the pulse phase is allowed to be weakly time-dependent $\dot{\phi}(t) \ll \omega_\text{d}$. The Rabi frequency is nonzero only during the short pulses. As an example, the results presented in Fig.~\ref{fig:signalReal}(a) are obtained by using box-shaped pulses, with $\Omega$ being a constant and nonzero only in the time intervals: $0 \leq t \leq 10~\text{\textmu s}$, $T + 10~\text{\textmu s} \leq t \leq T + 30~\text{\textmu s}$, and $2T + 30~\text{\textmu s} \leq t \leq 2T + 40~\text{\textmu s}$. 

For typical parameters, there is a separation of timescales between the internal dynamics of the atom and the external dynamics of its center-of-mass motion. This allows us to derive the approximate unitary time-evolution operators describing the action of the light pulses by neglecting the first line in the Hamiltonian~\eqref{eq:Hamiltonian}. Taking the instantaneous pulses and using the rotating wave approximation, these operators for perfect $\pi / 2$- and $\pi$-pulses are given by
\begin{align}
	U_\frac{\pi}{2} =& \frac{1}{\sqrt{2}} \mathbb{1} - \frac{\mathrm{i}}{\sqrt{2}} \left[\mathrm{e}^{\mathrm{i \phi(t)}}\ketbra{2}{1}  + \mathrm{e}^{\mathrm{-i \phi(t)}}\ketbra{1}{2} \right]\label{eq:piHalf}\\
	U_\pi =& -\mathrm{i} \left[\mathrm{e}^{\mathrm{i \phi(t)}}\ketbra{2}{1}  + \mathrm{e}^{\mathrm{-i \phi(t)}}\ketbra{1}{2}\right]\, ,\label{eq:pi}
\end{align}
where the pulse phase $\phi(t)$ is to be evaluated at the instant of the pulse ($0$, $T$, $2T$ in our case). Here we also assume that there is no detuning. 

Let the atom be prepared in the ground state of a harmonic trap and in the internal state $\ket{2}$, giving rise to the total state of the atom $\ket{\Psi(z,0)} = \psi(z,0) \ket{2}$. After the first $\pi / 2$-pulse the state reads $\ket{\Psi(z,0+\varepsilon)} = \psi(z,0) \big(\ket{2} - \mathrm{i}  \ket{1} \mathrm{e}^{\mathrm{-i \phi(0)}} \big) / \sqrt{2}$, where $\varepsilon$ is a infinitesimally small number. Due to the different potentials $V_{1}(z)$ and $V_{2}(z)$, the \emph{lower} $\psi_\text{l}(z,t)$ and \emph{upper} $\psi_\text{u}(z,t)$ wavepackets evolve independently, as shown in Fig.~\ref{fig:scheme1D}. The state immediately before the $\pi$-pulse therefore reads $\ket{\Psi(z,T-\varepsilon)} =  \big[\psi_\text{l}(z,T) \ket{2} - \mathrm{i} \psi_\text{u}(z,T) \ket{1} \mathrm{e}^{\mathrm{-i \phi(0)}} \big] / \sqrt{2}$, where the method to evaluate $\psi_\text{l}(z,T)$ and $\psi_\text{u}(z,T)$ is presented below. Directly after the $\pi$-pulse the state is given by $\ket{\Psi(z,T+\varepsilon)} =  -\mathrm{i} \big\{\psi_\text{l}(z,T) \ket{1} \mathrm{e}^{\mathrm{-i \phi(T)}} - \mathrm{i} \psi_\text{u}(z,T) \ket{2} \mathrm{e}^{\mathrm{-i [\phi(0) - \phi(T)]}} \big\} / \sqrt{2}$. The state after the last $\pi / 2$-pulse reads
\begin{equation}
	\begin{aligned}
		\ket{\Psi(z,2T)} =& - \frac{1}{2} \Big[\psi_\text{u}(z,2T) \mathrm{e}^{\mathrm{i} \Phi_{\text{P},1}} + \psi_\text{l}(z,2T) \mathrm{e}^{\mathrm{i} \Phi_{\text{P},2}} \Big] \ket{2}\\
		&+ \frac{\mathrm{i}}{2} \Big[\psi_\text{u}(z,2T) \mathrm{e}^{\mathrm{i} \Phi_{\text{P},3}} - \psi_\text{l}(z,2T) \mathrm{e}^{\mathrm{i} \Phi_{\text{P},4}} \Big] \ket{1}
	\end{aligned}
\end{equation}
with $\Phi_{\text{P},1} = \phi(T) - \phi(0)$, $\Phi_{\text{P},2} = \phi(2T) - \phi(T)$, $\Phi_{\text{P},3} = \phi(T) - \phi(0) - \phi (2T)$, $\Phi_{\text{P},4} = - \phi(T)$. The pulse phase contribution to the signal, Eq.~\eqref{eq:observable}, is $\Delta \Phi_\text{P} = \Phi_{\text{P},2} - \Phi_{\text{P},1} = \phi(0) - 2 \phi(T) + \phi(2T)$ and this can be made equal to $\pi / 2$ by choosing $\phi(t) = \pi t^2 / (4 T^2)$. For Fig.~\ref{fig:signalReal}(a) we have set $\phi(t) = 0$ to have no pulse phase contribution.

\noindent\textit{Wavepacket propagation.--}
A Gaussian wavepacket inside a harmonic potential keeps its Gaussian shape during the time evolution~\cite{Haas.2013}
\begin{equation}
	\psi(z,t) = \frac{1}{\sqrt{\sqrt{\pi} \sigma(t)}} \exp\left\{-\frac{[z - q(t)]^2}{2 \sigma^2(t)}\right\} \mathrm{e}^{\frac{\mathrm{i}}{\hbar} S(z,t)}
\end{equation}
with 
\begin{equation}
	S(z,t) = \frac{1}{2} a(t) [z - q(t)]^2 + b(t) [z - q(t)] + c(t) \,.
\end{equation}
Inserting this wavefunction into the 1D Schr\"odinger equation with the potential $V(z) = m \omega^2 z^2 / 2 - F z$ yields
\begin{align}
	&\ddot{\sigma}(t) + \omega^2 \sigma(t) = \frac{\hbar^2}{m^2 \sigma^3(t)}~\label{eq:Ermakov}\\
	&\ddot{q}(t) + \omega^2 q(t) = \frac{F}{m}\label{eq:Newton}\\
	&a(t) = m \frac{\dot{\sigma}(t)}{\sigma(t)}\\
	&b(t) = m \dot{q}(t)\\
	&\dot{c}(t) = \frac{1}{2} m \dot{q}^2(t) - \frac{1}{2} m \omega^2 q^2(t) + F q(t) - \frac{\hbar^2}{2m \sigma^2(t)}\,.
\end{align}
To propagate the wavepackets in their traps during the two intervals of the MZI, we solve the Ermakov equation~\eqref{eq:Ermakov} and the Newton equation~\eqref{eq:Newton} with the corresponding trap parameters and initial conditions. The contrast and phase of the 1D MZI $C\mathrm{e}^{\mathrm{i}\varphi} = \int \mathrm{d}z \psi_\text{u}^\ast(z,2T) \psi_\text{l}(z,2T)$ then depend on the final value of the solutions $q_\text{u}(2T)$, $q_\text{l}(2T)$, $\sigma_\text{u}(2T)$, and $\sigma_\text{l}(2T)$. The generalization to the 3D MZI is straightforward by using product wavefunctions $\psi(\mathbf{x},t) = \prod_i \psi_i(x_i,t)$. We provide more details on the calculation of $C$ and $\varphi$ in the Supplemental Material~\cite{Note1}.

\noindent\textit{Optical dipole trap model.--}
An alkali atom inside a far-detuned, linearly polarized, monochromatic Gaussian laser beam, with its focus being at the origin of the coordinate system and propagating along the $x_i$-axis with the intensity
\begin{equation}
	I_i(\mathbf{x}) = \frac{2 P_i}{\pi w_i^2 (1 + x_i^2 / R_i^2)} \exp\left[-\frac{2(\mathbf{x}^2 - x_i^2)}{w_i^2 (1 + x_i^2 / R_i^2)}\right]\,,
\end{equation}
is affected by the optical dipole potential $V_i(\mathbf{x}) = \hbar \kappa I_i(\mathbf{x}) / 8$ with $\kappa = (\kappa_\text{D1} + 2 \kappa_\text{D2}) / 3$ and
\begin{equation}
	\kappa_\text{D} = \frac{12 \pi c^2}{\hbar \omega_\text{D}^3} \left(\frac{\Gamma_\text{D}}{\omega_i - \omega_\text{D}} - \frac{\Gamma_\text{D}}{\omega_i + \omega_\text{D}}\right)\,,
\end{equation}
where $P_i$ is the power, $w_i$ is the beam waist, $R_i = \pi w_i^2 / \lambda_i$ is the Rayleigh range, $\lambda_i$ is the wavelength of the laser, $\omega_i = 2 \pi c / \lambda_i$ is its frequency, $\omega_\text{D}$ is the transition frequency of the $\text{D}$-lines $\text{D1}$ and $\text{D2}$, and $\Gamma_\text{D}$ is the corresponding linewidth~\cite{Grimm.2000}.

In the main text we consider a crossed optical dipole trap where two Gaussian laser beams with the same power $P$, beam waist $w$, and wavelength $\lambda$ have their focus at the origin of our coordinate system. One laser beam propagates along the $x$-axis and the other one along the $y$-axis. The total optical dipole potential is therefore given by $V_\text{OD}(\mathbf{x}) = V_x(\mathbf{x}) + V_y(\mathbf{x})$. The wavelength of $\lambda = 1064~\text{nm}$ is red-detuned with respect to the $\text{D}$-lines of $^{87}$Rb~\cite{Steck.2003} which gives rise to $\kappa < 0$. Consequently, the atoms are high-field seeking and trapped in the focus point where the intensity is largest.

In the presence of the linear gravitational potential $V_g(z) = mgz$ with $g > 0$, the atoms are sagged downwards out of the focus. The minimum of the total potential $V(\mathbf{x}) = V_\text{OD}(\mathbf{x}) + V_g(z)$ is now located at $(0,0,z_0)$ with 
\begin{equation}
	z_0 = - \frac{1}{2} w \sqrt{-W\left(-\frac{g^2 m^2 \pi^2 w^6}{\hbar^2 \kappa^2 P^2}\right)}\,,
\end{equation}
where $W(x)$ is the Lambert function. Note that the minimum vanishes at $g^2 m^2 \pi^2 w^6 / (\hbar^2 \kappa^2 P^2) \geq 1 / \mathrm{e}$. Expanding the potential up to the second order at this point yields the trap frequencies $\omega_{x,y} = \sqrt{-(2 R^2 + w^2 - 2 z_0^2) \hbar \kappa P / (2 \pi m w^4 R^2)} \exp\left\{- z_0^2 / w^2\right\}$ and
\begin{equation}\label{eq:omegaDipole}
	\omega_z = \sqrt{- 2 \frac{w^2 - 4 z_0^2}{\pi m w^6} \hbar \kappa P} \exp\left\{- \frac{z_0^2}{w^2}\right\}\,.
\end{equation}
Expanding the potential up to higher orders gives the anharmonic terms $\alpha (z - z_0)^3$ and $\beta (z - z_0)^4$ with 
\begin{align}
	\alpha &= 4 \frac{3 w^2 - 4 z_0^2}{3 \pi w^8}  z_0 \hbar \kappa P \exp\left\{- 2 \frac{z_0^2}{w^2}\right\}\\
	\beta &= \frac{3 w^4 - 24 w^2 z_0^2 + 16 z_0^4}{3 \pi w^{10}} \hbar \kappa P \exp\left\{- 2 \frac{z_0^2}{w^2}\right\}\,.\label{eq:anharmonicDipole}
\end{align}

\noindent\textit{Anharmonic oscillator.--}
The period of a classical 1D finite motion in a potential $V(z)$ is given by~\cite{Landau.1976}
\begin{equation}\label{eq:anharmonic}
	\mathcal{T}(E) = \sqrt{2 m} \int_{z_1(E)}^{z_2(E)} \frac{dz}{\sqrt{E - V(z)}}
\end{equation}
where $E$ is the energy and the integration takes place between the classical turning points $z_1(E)$ and $z_2(E)$. Inserting the potential $V(z) = m \omega_z^2 z^2 / 2 + \alpha z^3 + \beta z^4$ and calculating the integral perturbatively for $|\alpha| \ll 1$ and $|\beta| \ll 1$, we obtain the asymptotic expansion
\begin{equation}\label{eq:anharmonicApprox}
	\omega \approx \omega_z + \omega_z \left(\frac{3}{2} \beta - \frac{15}{4} \frac{\alpha^2}{m \omega_z^2}\right) \frac{Z_0^2}{m \omega_z^2}
\end{equation}
for the frequency $\omega = 2 \pi / \mathcal{T}$. Here, $Z_0 = \sqrt{2 E / (m \omega_z^2)}$ is the classical turning point of the purely harmonic motion.

We use Eq.~\eqref{eq:anharmonicApprox} to motivate the fit models for the data shown in Fig.~\ref{fig:signalReal}(b) where we plot the frequencies against the distance $\Delta z = |z_{0,1} - z_{0,2}|$ between the trap centers. Within the order we are interested in, we can set $Z_0 = \Delta z$ in Eq.~\eqref{eq:anharmonicApprox}. The data was obtained by moving the minimum $z_{0,2}$ while keeping $z_{0,1} \approx -13.72~\text{\textmu m}$ constant. To motivate the model for fitting $\omega_{2,z}$ we therefore express the former as $z_{0,2}(\Delta z) = z_{0,1} + \Delta z$ and insert this into Eqs.~\eqref{eq:omegaDipole}-\eqref{eq:anharmonicDipole}. The results are used as $\omega_z$, $\alpha$, and $\beta$ in Eq.~\eqref{eq:anharmonicApprox} which leads to a dependence of these quantities on $\Delta z$, {\it i.e.} we have $\omega_z(\Delta z)$, $\alpha(\Delta z)$, and $\beta(\Delta z)$. Expanding Eq.~\eqref{eq:anharmonicApprox} up to the second order in $\Delta z$ provides a good approximation for our parameters and gives rise to a contribution that is linear in $\Delta z$. Consequently, we employ the fit model $\omega_{2,z} = k_1 + k_2 \Delta z + k_3 \Delta z^2$.

For fitting the frequency $\omega_{1,z}$ we have to come back to the trajectories of the wavepackets. When determining the frequency by starting the interferometer in state $\ket{1}$, the wavepackets move extensively in both traps. For this reason we fit to same model as for $\omega_{2,z}$. Starting the interferometer in state $\ket{2}$ means that at $T = n \mathcal{T}_1$ the wavepackets only move in the trap of state $\ket{1}$. In this case the fit model reads $\omega_{1,z} = k_1 + k_3 \Delta z^2$ in direct analogy to Eq.~\eqref{eq:anharmonicApprox}. Moreover, the coefficient $k_3$ of this last fit can be used to obtain the upper bounds $\alpha^2 < 4 m^2 \omega_z^3 |k_3| / 15$ and $|\beta| < 2 m \omega_z |k_3| / 3$ on the magnitudes $\alpha$ and $\beta$ of the anharmonic terms. Note that this only works if the fit results in a negative $k_3$ and it is known from earlier modeling that $\beta < 0$.
	\clearpage
	\setcounter{page}{1}
\setcounter{equation}{0}
\renewcommand{\theequation}{S.\arabic{equation}}
\setcounter{figure}{0}
\renewcommand{\thefigure}{S\arabic{figure}}
\setcounter{table}{0}
\renewcommand{\thetable}{S\arabic{table}}
\onecolumngrid
\begin{center}
	{\normalfont\large\bfseries\centering {\it Supplemental material for}\\ 
	Mach-Zehnder interferometer for in-situ characterization of atom traps}\\[23pt]	
\end{center}
\onecolumngrid
\section{Wavepacket propagation}
A Gaussian wavepacket propagating inside a harmonic potential stays Gaussian. Assuming that any interferometer starts with a Gaussian, most likely the ground state of either harmonic trap, it is therefore easier (and exact) to propagate the first and second moment of the Gaussian, instead of using the full path integral propagator. We use the well-known ansatz wavefunction~\cite{Haas.2013}
\begin{equation}\label{eq:ansatz}
	\psi(z,t) = \frac{1}{\sqrt{\sqrt{\pi} \sigma(t)}} \exp\left\{-\frac{[z - q(t)]^2}{2 \sigma^2(t)}\right\} \mathrm{e}^{\frac{\mathrm{i}}{\hbar} \left\{\frac{1}{2} a(t) [z - q(t)]^2 + b(t) [z - q(t)] + c(t)\right\}}
\end{equation}
which after inserting into the time-dependent Schrödinger equation
\begin{equation}
    \mathrm{i}\hbar \frac{\partial}{\partial t} \psi(z,t) = \left[-\frac{\hbar^2}{2m} \frac{\partial^2}{\partial z^2} + \frac{1}{2} m \omega^2 z^2 - F z\right] \psi(z,t)
\end{equation}
and separating real and imaginary part for each order in $z$ yields
\begin{align}
    a(t) &= m \frac{\dot{\sigma}(t)}{\sigma(t)} \label{eq:alpha}\\
    \ddot{\sigma}(t) + \omega^2 \sigma(t) &= \frac{\hbar^2}{m^2 \sigma^3(t)}\\
    b(t) &= m \dot{q}(t)\\
    \ddot{q}(t) + \omega^2 q(t) &= \frac{F}{m}\\
    \dot{c}(t) & = \frac{1}{2} m \dot{q}^2(t) - \frac{1}{2} m \omega^2 q^2(t) + F q(t) - \frac{\hbar^2}{2m \sigma^2(t)}\,.\label{eq:gamma}
\end{align}
The solution for $q(t)$ and $\sigma(t)$ in terms of their initial values $q_0$, $\dot{q}_0$ and $\sigma_0$, $\dot{\sigma}_0$ is
\begin{align}
    q(t) &= \frac{F}{m \omega^2} + \left(q_0 - \frac{F}{m \omega^2}\right) \cos(\omega t) + \frac{\dot{q}_0}{\omega} \sin(\omega t) \label{eq:qSol}\\
    \sigma(t) &= \sigma_0 \sqrt{\frac{1}{2} \left(1 + \frac{\dot{\sigma}_0^2}{\omega^2 \sigma_0^2} + \frac{\hbar^2}{m^2 \omega^2 \sigma_0^4}\right) + \frac{1}{2} \left(1 - \frac{\dot{\sigma}_0^2}{\omega^2 \sigma_0^2} - \frac{\hbar^2}{m^2 \omega^2 \sigma_0^4}\right) \cos(2 \omega t) + \frac{\dot{\sigma}_0}{\omega \sigma_0} \sin(2 \omega t)} \label{eq:sigmaSol}
\end{align}
which determines $a(t)$ and $b(t)$ as well as
\begin{equation}\label{eq:gammaSol}
    c(t) = \int_0^t \mathrm{d}t^\prime \ \left[\frac{1}{2} m \dot{q}^2(t^\prime) - \frac{1}{2} m \omega^2 q^2(t^\prime) + F q(t^\prime) - \frac{\hbar^2}{2 m \sigma^2(t^\prime)}\right] + c_0
\end{equation}
up to some initial value $c_0$. We can set $c_0 = 0$ since any interferometer starts from a single wavepacket that is split apart by pulses. Consequently, all wavepackets share the same $c_0$ which therefore turns into a global phase.

The action $c(t)$ contains a contribution from the classical center-of-mass (CM) motion
\begin{equation}
    c_\text{CM}(t) = \int_0^t \mathrm{d}t^\prime \left[\frac{1}{2} m \dot{q}^2(t^\prime) - \frac{1}{2} m \omega^2 q^2(t^\prime) + F q(t^\prime)\right]
\end{equation}
and breathing of the wavepacket
\begin{equation}
    c_\text{B}(t) = - \frac{\hbar^2}{2 m}\int_0^t \mathrm{d}t^\prime \ \frac{1}{\sigma^2(t)}\,,
\end{equation}
respectively. For later reference we evaluate these integrals using the general solutions~\eqref{eq:qSol} and~\eqref{eq:sigmaSol}. The CM part simplifies by means of partial integration and insertion of~\eqref{eq:qSol}
\begin{equation}\label{eq:cmAction}
    \begin{aligned}
        c_\text{CM}(t) &= \frac{1}{2} m \dot{q}(t^\prime) q(t^\prime)\Big|_0^t + \frac{1}{2} F \int_0^t \mathrm{d}t^\prime q(t^\prime)\\
        &= \frac{1}{2} m \left[\dot{q}(t) q(t) - \dot{q}_0 q_0\right] + \frac{F^2}{2m\omega^2} t + \frac{F}{2 \omega} \left(q_0 - \frac{F}{m \omega^2}\right) \sin(\omega t) - \frac{F \dot{q}_0}{2 \omega^2} \left[\cos(\omega t) - 1\right]
    \end{aligned} 
\end{equation}
and the breathing part is
\begin{equation}\label{eq:breathingAction}
    c_\text{B}(t) = - \frac{\hbar}{2} \left\{\arctan\left[\frac{m \dot{\sigma}_0 \sigma_0}{\hbar} + \left(\frac{m \dot{\sigma}_0^2}{\hbar \omega} + \frac{\hbar}{m \omega \sigma_0^2}\right) \tan(\omega t)\right] - \arctan\left(\frac{m \dot{\sigma}_0 \sigma_0}{\hbar}\right) + \pi \left\lfloor \frac{1}{2} + \frac{\omega t}{\pi} \right\rfloor\right\} 
\end{equation}
with the floor function $\lfloor x \rfloor$.

Later on we also require the energy which is the expectation value of the Hamiltonian with respect to the wavefunction~\eqref{eq:ansatz}
\begin{equation}
    \begin{aligned}
        E &= \frac{b^2(t)}{2m} + \frac{1}{2} m \omega^2 q^2(t) - F q(t) + \frac{a^2(t) \sigma^2(t)}{4m} + \frac{\hbar^2}{4 m \sigma^2(t)} + \frac{1}{4} m \omega^2 \sigma^2(t)\\
        &= \frac{1}{2} m \dot{q}^2(t) + \frac{1}{2} m \omega^2 q^2(t) - F q(t) + \frac{1}{4} m \dot{\sigma}^2(t) + \frac{\hbar^2}{4 m \sigma^2(t)} + \frac{1}{4} m \omega^2 \sigma^2(t)
    \end{aligned}
\end{equation}
and splits into a CM and breathing part
\begin{align}
    E_\text{CM} &= \frac{1}{2} m \dot{q}^2(t) + \frac{1}{2} m \omega^2 q^2(t) - F q(t) = \text{const.}\\
    E_\text{B} &= \frac{1}{4} m \dot{\sigma}^2(t) + \frac{\hbar^2}{4 m \sigma^2(t)} + \frac{1}{4} m \omega^2 \sigma^2(t) =\text{const.} \label{eq:breathingEnergy}
\end{align}
which are conserved separately within each interval of an interferometer.

Finally, we require the overlap of two Gaussian wavepackets $\psi_\text{u}(z,t)$ and $\psi_\text{l}(z,t)$ from which we can extract the contrast and phase of an interferometer later on. Completing the square in Eq.~\eqref{eq:ansatz} yields
\begin{equation}
    \psi(z,t) = \frac{1}{\sqrt{\sqrt{\pi} \sigma(t)}} \exp\left\{-\frac{1}{2} W(t) \left[z - Q(t)\right]^2 - \frac{b^2(t)}{2\hbar^2 W(t)} + \frac{\mathrm{i}}{\hbar} c(t)\right\}
\end{equation}
with the complex quantities
\begin{align}
    W(t) &= \frac{1}{\sigma^2(t)} - \frac{\mathrm{i}}{\hbar} a(t)\\
    Q(t) &= q(t) + \frac{\mathrm{i}}{\hbar} \frac{b(t)}{W(t)}
\end{align}
and repeating this process for the overlap
\begin{equation}
    \begin{aligned}
        \psi_\text{u}^\ast(z,t) \psi_\text{l}(z,t) =& \ \frac{1}{\sqrt{\pi \sigma_\text{u}(t) \sigma_\text{l}(t)}} \exp\Bigg\{-\frac{1}{2} \big[W_\text{l}(t) + W_\text{u}^\ast(t)\big] \left[z -\frac{Q_\text{l}(t) W_\text{l}(t) + Q_\text{u}^\ast(t) W_\text{u}^\ast(t)}{W_\text{l}(t) + W_\text{u}^\ast(t)}\right]^2\\
        &-\frac{1}{2} \frac{W_\text{l}(t) W_\text{u}^\ast(t)}{W_\text{l}(t) + W_\text{u}^\ast(t)} \left[Q_\text{l}(t) - Q_\text{u}^\ast(t)\right]^2 - \frac{b_\text{l}^2(t)}{2\hbar^2 W_\text{l}(t)} - \frac{b_\text{u}^2(t)}{2\hbar^2 W_\text{u}^\ast(t)} + \frac{\mathrm{i}}{\hbar}\left[c_\text{l}(t) - c_\text{u}(t)\right]\Bigg\}
    \end{aligned}
\end{equation}
results in 
\begin{equation}\label{eq:overlap}
    \begin{aligned}
        \int_{-\infty}^\infty \mathrm{d}z \ \psi_\text{u}^\ast(z,t) \psi_\text{l}(z,t) = & \ \sqrt{\frac{2}{\sigma_\text{u}(t) \sigma_\text{l}(t)}} \frac{1}{\sqrt{W_\text{l}(t) + W_\text{u}^\ast(t)}} \exp\Bigg\{-\frac{1}{2} \frac{W_\text{l}(t) W_\text{u}^\ast(t)}{W_\text{l}(t) + W_\text{u}^\ast(t)} \left[Q_\text{l}(t) - Q_\text{u}^\ast(t)\right]^2\\
        &- \frac{b_\text{l}^2(t)}{2\hbar^2 W_\text{l}(t)} - \frac{b_\text{u}^2(t)}{2\hbar^2 W_\text{u}^\ast(t)} + \frac{\mathrm{i}}{\hbar}\left[c_\text{l}(t) - c_\text{u}(t)\right]\Bigg\}\,.
    \end{aligned}
\end{equation}
Since the real part of $W(t)$ is strictly positive, the complex prefactor $1/\sqrt{W_\text{l}(t) + W_\text{u}^\ast(t)}$ is easily split into modulus and phase
\begin{equation}
    \frac{1}{\sqrt{W_\text{l}(t) + W_\text{u}^\ast(t)}} = \frac{1}{\sqrt{\left|W^\ast_\text{l}(t) + W_\text{u}(t)\right|}} \exp\left\{\mathrm{i} \frac{1}{2} \arctan\left[\frac{\operatorname{Im}\left[W^\ast_\text{l}(t) + W_\text{u}(t)\right]}{\operatorname{Re}\left[W^\ast_\text{l}(t) + W_\text{u}(t)\right]}\right] \right\}
\end{equation}
and we can give formal expressions for the contrast $C(t) = C_\text{B}(t)  C_\text{CM}(t)$ and phase $\varphi(t) = \varphi_\text{B}(t) +  \varphi_\text{CM}(t)$
\begin{align}
    C_\text{B}(t) =& \sqrt{\frac{2}{\sigma_\text{u}(t) \sigma_\text{l}(t)}} \frac{1}{\sqrt{\left|W^\ast_\text{l}(t) + W_\text{u}(t)\right|}}\label{eq:contrasBreathing}\\
    C_\text{CM}(t) =& \exp\left\{\operatorname{Re}\left[-\frac{1}{2} \frac{W_\text{l}(t) W_\text{u}^\ast(t)}{W_\text{l}(t) + W_\text{u}^\ast(t)} \left[Q_\text{l}(t) - Q_\text{u}^\ast(t)\right]^2 - \frac{b_\text{l}^2(t)}{2\hbar^2 W_\text{l}(t)} - \frac{b_\text{u}^2(t)}{2\hbar^2 W_\text{u}^\ast(t)}\right]\right\}\\
    \varphi_\text{B}(t) =& \frac{1}{2}\arctan\left[\frac{\operatorname{Im}\left[W^\ast_\text{l}(t) + W_\text{u}(t)\right]}{\operatorname{Re}\left[W^\ast_\text{l}(t) + W_\text{u}(t)\right]}\right] + \frac{c_\text{B,l}(t) - c_\text{B,u}(t)}{\hbar}\label{eq:phaseBreathing}\\
    \varphi_\text{CM}(t) =& \exp\Bigg\{\operatorname{Im}\left[-\frac{1}{2} \frac{W_\text{l}(t) W_\text{u}^\ast(t)}{W_\text{l}(t) + W_\text{u}^\ast(t)} \left[Q_\text{l}(t) - Q_\text{u}^\ast(t)\right]^2 - \frac{b_\text{l}^2(t)}{2\hbar^2 W_\text{l}(t)} - \frac{b_\text{u}^2(t)}{2\hbar^2 W_\text{u}^\ast(t)}\right]\\
    &\phantom{\exp\Bigg\{} + \frac{c_\text{CM,l}(t) - c_\text{CM,u}(t)}{\hbar}\bigg\}
\end{align}
in terms of breathing and CM contributions.
\section{Application to a Mach-Zehnder Interferometer}
When the pulses in an interferometer are approximated as an instantaneous population transfer, the above results can be used to calculate the contrast and phase of any interferometer by making $\omega$ and $F$ piecewise time-dependent. Equations~\eqref{eq:alpha}-\eqref{eq:gamma} stay unchanged but the solutions~\eqref{eq:qSol} and~\eqref{eq:sigmaSol} have to be applied with new initial conditions in each time interval. Likewise the action becomes a sum over contributions form each interval. The respective integrals can, however, be reduced to the ones in Eqs.~\eqref{eq:cmAction} and~\eqref{eq:breathingAction} by substituting the integration variable.

We apply this scheme to the MZI with ideal pulses in Fig.~\ref{fig:scheme1D} which results in the two-path-interference of an \emph{upper} and \emph{lower} wavepacket with
\begin{align}
    \omega_\text{u}(t) &= 
    \begin{cases} 
        \omega_1 &  0 < t \leq T \\
        \omega_2 &  T < t \leq 2T
   \end{cases} \qquad F_\text{u}(t) = 
    \begin{cases} 
        F_1 &  0 < t \leq T \\
        F_2 &  T < t \leq 2T
   \end{cases}\\
   \omega_\text{l}(t) &= 
    \begin{cases} 
        \omega_2 &  0 < t \leq T \\
        \omega_1 &  T < t \leq 2T
   \end{cases} \qquad F_\text{l}(t) = 
    \begin{cases} 
        F_2 &  0 < t \leq T \\
        F_1 &  T < t \leq 2T \,.
   \end{cases}
\end{align}
The interferometer starts in the ground state of the trap in state $\ket{2}$ which means the initial values for the first time interval are
\begin{align}
    q_{0,\text{u}}^{(1)} &= q_{0,\text{l}}^{(1)} = \frac{F_2}{m \omega_2^2}\\
    \dot{q}_{0,\text{u}}^{(1)} &= \dot{q}_{0,\text{l}}^{(1)} = 0\\
    \sigma_{0,\text{u}}^{(1)} &= \sigma_{0,\text{l}}^{(1)} = \sqrt{\frac{\hbar}{m \omega_2}}\\
    \dot{\sigma}_{0,\text{u}}^{(1)} &= \dot{\sigma}_{0,\text{l}}^{(1)} = 0
\end{align}
and the corresponding solutions are ($0 < t \leq T$)
\begin{align}
    q_\text{u}^{(1)}(t) &= \frac{F_1}{m \omega_1^2} + \left(\frac{F_2}{m \omega_2^2} - \frac{F_1}{m \omega_1^2}\right) \cos(\omega_1 t)\\
    q_\text{l}^{(1)}(t) &= \frac{F_2}{m \omega_2^2}\\
    \sigma_\text{u}^{(1)}(t) &= \sqrt{\frac{\hbar}{m \omega_2}} \sqrt{\frac{1}{2} \left(1 + \frac{\omega_2^2}{\omega_1^2}\right) + \frac{1}{2} \left(1 - \frac{\omega_2^2}{\omega_1^2}\right) \cos(2 \omega_1 t)}\\
    \sigma_\text{l}^{(1)}(t) &= \sqrt{\frac{\hbar}{m \omega_2}}\,.
\end{align}

In the second time interval the initial conditions are given by the end points of the first interval. At this point it is important to note that in the second interval the initial values of the upper wavepacket  are the final values of the lower wavepacket, that is
\begin{align}
    q_\text{l}^{(2)}(2T) &= q_\text{u}^{(2)}(T)\\
    \dot{q}_\text{l}^{(2)}(2T) &= \dot{q}_\text{u}^{(2)}(T) \\
    \sigma_\text{l}^{(2)}(2T) &= \sigma_\text{u}^{(2)}(T) \\
    \dot{\sigma}_\text{l}^{(2)}(2T) &= \dot{\sigma}_\text{u}^{(2)}(T)\,,
\end{align}
which means they later enter into the contrast $C(2T)$ and phase $\varphi(2T)$. For this reason we dignify the initial values of the upper wavepacket in the second interval with a special shortened notation and write ($T < t \leq 2T$)
\begin{align}
    q_\text{u}^{(2)}(t) &= \frac{F_2}{m \omega_2^2} + \left(q_0 - \frac{F_2}{m \omega_2^2}\right) \cos\left[\omega_2 (T - t)\right] + \frac{\dot{q}_0}{\omega_2} \sin\left[\omega_2 (T - t)\right]\label{eq:qU2}\\
    \sigma_\text{u}^{(2)}(t) &= \sigma_0 \sqrt{\frac{1}{2} \left(1 + \frac{\dot{\sigma}_0^2}{\omega_2^2 \sigma_0^2} + \frac{\hbar^2}{m^2 \omega_2^2 \sigma_0^4}\right) + \frac{1}{2} \left(1 - \frac{\dot{\sigma}_0^2}{\omega_2^2 \sigma_0^2} - \frac{\hbar^2}{m^2 \omega_2^2 \sigma_0^4}\right) \cos\left[2 \omega_2 (T - t)\right] + \frac{\dot{\sigma}_0}{\omega_2 \sigma_0} \sin\left[2 \omega_2 (T - t)\right]}\label{eq:sigmaU2}\\
\end{align}
(note the shifted time arguments) with
\begin{align}
    q_0 &= \frac{F_1}{m \omega_1^2} + \left(\frac{F_2}{m \omega_2^2} - \frac{F_1}{m \omega_1^2}\right) \cos(\omega_1 T)\\
    \dot{q}_0 &= -\omega_1 \left(\frac{F_2}{m \omega_2^2} - \frac{F_1}{m \omega_1^2}\right) \sin(\omega_1 T)\\
    \sigma_0 &= \sqrt{\frac{\hbar}{m \omega_2}} \sqrt{\frac{1}{2} \left(1 + \frac{\omega_2^2}{\omega_1^2}\right) + \frac{1}{2} \left(1 - \frac{\omega_2^2}{\omega_1^2}\right) \cos(2 \omega_1 T)}\label{eq:sigma0U2}\\
    \dot{\sigma}_0 &= -\frac{1}{2} \sqrt{\frac{\hbar}{m \omega_2}} \frac{\omega_1 \left(1 - \frac{\omega_2^2}{\omega_1^2}\right) \sin(2 \omega_1 T)}{\sqrt{\frac{1}{2} \left(1 + \frac{\omega_2^2}{\omega_1^2}\right) + \frac{1}{2} \left(1 - \frac{\omega_2^2}{\omega_1^2}\right) \cos(2 \omega_1 T)}}\label{eq:dsigma0U2}\,.
\end{align}
For the lower wavepacket we have the same initial conditions as in the first interval and therefore ($T < t \leq 2T$)
\begin{align}
    q_\text{l}^{(2)}(t) &= \frac{F_1}{m \omega_1^2} + \left(\frac{F_2}{m \omega_2^2} - \frac{F_1}{m \omega_1^2}\right) \cos\left[\omega_1 (T - t)\right]\\
    \sigma_\text{l}^{(2)}(t) &= \sqrt{\frac{\hbar}{m \omega_2}} \sqrt{\frac{1}{2} \left(1 + \frac{\omega_2^2}{\omega_1^2}\right) + \frac{1}{2} \left(1 - \frac{\omega_2^2}{\omega_1^2}\right) \cos\left[2 \omega_1 (T - t)\right]}\,.
\end{align}
\subsection{Breathing contribution}
Significant simplifications of the formulas can be achieved for the contrast~\eqref{eq:contrasBreathing} and phase~\eqref{eq:phaseBreathing} of the breathing motion. We begin by writing down the $c_\text{B}$ contribution to the phase which splits into four terms, one from each wavepacket and interval, that can all be obtained from Eq.~\eqref{eq:breathingAction} by making the suitable replacements for $\omega$ and the initial conditions as well as setting $t = T$. The floor functions cancel out in the process of combining the four terms and we have
\begin{equation}
    \frac{c_\text{B,l}(2T) - c_\text{B,u}(2T)}{\hbar} = -\frac{1}{2} \left[\arctan(\tan(\omega_2 T)) + \arctan\left(\frac{m \dot{\sigma}_0 \sigma_0}{\hbar}\right) - \arctan\left[\frac{m \dot{\sigma}_0 \sigma_0}{\hbar} + \left(\frac{m \dot{\sigma}_0^2}{\hbar \omega_2} + \frac{\hbar}{m \omega_2 \sigma_0^2}\right) \tan(\omega_2 t)\right]\right]
\end{equation}
which can be reduce by applying the angle sum identities for the arctangent
\begin{equation}
    \frac{c_\text{B,l}(2T) - c_\text{B,u}(2T)}{\hbar} = -\frac{1}{2} \arctan\left[\frac{\sigma_0^2 - \frac{\hbar}{m \omega_2} + \frac{\dot{\sigma}_0 \sigma_0}{\omega_2} \tan(\omega_2 T)}{\sigma_0^2 + \frac{\hbar}{m \omega_2} \tan^2(\omega_2 T) + \frac{\dot{\sigma}_0 \sigma_0}{\omega_2} \tan(\omega_2 T)} \tan(\omega_2 T)\right]\,.
\end{equation}

The first term in the breathing phase~\eqref{eq:phaseBreathing} reads
\begin{equation}
    \frac{1}{2}\arctan\left[\frac{\operatorname{Im}\left[W^\ast_\text{l}(2T) + W_\text{u}(2T)\right]}{\operatorname{Re}\left[W^\ast_\text{l}(2T) + W_\text{u}(2T)\right]}\right] = \frac{1}{2}\arctan\left[\frac{m}{\hbar} \sigma_0 \sigma_\text{u}^{(2)}(2T) \frac{\dot{\sigma_0}\sigma_\text{u}^{(2)}(2T) - \dot{\sigma}_\text{u}^{(2)}(2T) \sigma_0}{\sigma_0^2 + \left(\sigma_\text{u}^{(2)}(2T)\right)^2}\right]
\end{equation}
where we have already used the fact that $\sigma_\text{l}^{(2)}(2T) = \sigma_0$ and $\dot{\sigma}_\text{l}^{(2)}(2T) = \dot{\sigma}_0$. Inserting Eq.~\eqref{eq:sigmaU2} and again using the angle sum identities yields
\begin{equation}\label{eq:phaseBreathing2}
    \varphi_\text{B}(2T) = \frac{1}{2} \arctan\left\{\frac{\left[\frac{\dot{\sigma}_0^2 \sigma_0^2}{\omega_2^2} + \left(\sigma_0^2 - \frac{\hbar}{m \omega_2}\right)^2\right] \sin(2 \omega_2 T)}{\frac{\dot{\sigma}_0^2 \sigma_0^2}{\omega_2^2} +  \left(\sigma_0^2 + \frac{\hbar}{m \omega_2}\right)^2 - \left[\frac{\dot{\sigma}_0^2 \sigma_0^2}{\omega_2^2} + \left(\sigma_0^2 - \frac{\hbar}{m \omega_2}\right)^2\right] \cos(2 \omega_2 T)}\right\}
\end{equation}
which depends on $\dot{\sigma}_0$ only as $\dot{\sigma}_0^2$. The final simplification comes about by using the energy relation~\eqref{eq:breathingEnergy} for the upper wavepacket in the second interval which we denote as
\begin{equation}\label{eq:breathingEnergyU2}
     E_\text{B} = \frac{1}{4} m \dot{\sigma}_0^2 + \frac{\hbar^2}{4 m \sigma_0^2} + \frac{1}{4} m \omega_2^2 \sigma_0^2 = \text{const.}
\end{equation}
and solve for $\dot{\sigma}_0^2$
\begin{equation}\label{eq:dsigmaU2}
      \dot{\sigma}_0^2 = \frac{4}{m} \left[\hbar \omega_2 \overline{E}_\text{B} - \frac{\hbar^2}{4 m \sigma_0^2} - \frac{1}{4} m \omega_2^2 \sigma_0^2\right]
\end{equation}
with the scaled energy
\begin{equation}
      \overline{E}_\text{B} = \frac{E_\text{B}}{\hbar \omega_2} = \frac{(\omega_1^2 + \omega_2^2)^2 - (\omega_1^2 - \omega_2^2)^2 \cos(2 \omega_1 T)}{8 \omega_1^2 \omega_2^2}
\end{equation}
calculated by inserting Eqs.~\eqref{eq:sigma0U2} and~\eqref{eq:dsigma0U2} into Eq.~\eqref{eq:breathingEnergyU2}. Upon inserting Eq.~\eqref{eq:dsigmaU2} into Eq.~\eqref{eq:phaseBreathing2} we obtain the compact expression
\begin{equation}\label{eq:breathingPhaseFinal}
    \varphi_\text{B}(2T) = \frac{1}{2} \arctan\left[\frac{(2 \overline{E}_\text{B} - 1) \sin(2 \omega_2 T)}{2 \overline{E}_\text{B} + 1 - (2 \overline{E}_\text{B} - 1) \cos(2 \omega_2 T)}\right]
\end{equation}
and a similar derivation yields
\begin{equation}\label{eq:breathingContrastFinal}
    C_\text{B}(2T) = \left[\frac{2}{1 + 4 \overline{E}_\text{B}^2 + (1 - 4 \overline{E}_\text{B}^2) \cos(2 \omega_2 T)}\right]^{1/4}\,.
\end{equation}
These results are useful because they are the only contributions for the transversal oscillators when the force is aligned with one of the principal axes of a 3D trap. In case of the discussion in the main text we have $C_x(2T) = C_{\text{B},x}(2T)$, $C_y(2T) = C_{\text{B},y}(2T)$, $\varphi_x(2T) = \varphi_{\text{B},x}(2T)$, and $\varphi_y(2T) = \varphi_{\text{B},y}(2T)$. 

Note that if $\omega_1 = \omega_2$ then $\overline{E}_\text{B} = 1 / 2$, $\varphi_\text{B}(2T) = 0$, and $C_\text{B}(2T) = 1$ as expected. The same holds true if $\omega_1 \neq \omega_2$ but $T = \pi n / \omega_1$ with integer $n$. If $T = \pi n / \omega_2$ with integer $n$, then $\varphi_\text{B}(2T) = 0$ and $C_\text{B}(2T) = 1$ no matter the value of $\overline{E}_\text{B}$.
\subsection{Center-of-mass contribution}
Similar simplifications for the CM contribution to the contrast and phase have so far eluded us, because they are a more complicated mixture of the classical motion $q(t)$ and breathing $\sigma(t)$. At this point we resort to simply inserting Eqs.~\eqref{eq:qU2}-\eqref{eq:dsigma0U2} into $C_\text{CM}(2T)$ and $\varphi_\text{CM}(2T)$ whenever we plot results. Note that $q_\text{l}^{(2)}(2T) = q_0$ and $\dot{q}_\text{l}^{(2)}(2T) = \dot{q}_0$ as was the case for the breathing motion as well.

The only straightforward simplification results for the difference of the action which becomes
\begin{equation}
    \frac{c_\text{CM,l}(2T) - c_\text{CM,u}(2T)}{\hbar} = \frac{m}{2 \hbar \omega_2} \left[(\tilde{q}_0^2 \omega_2^2 - \dot{q}_0^2) \cos(\omega_2 T) + 2 \dot{q}_0 \tilde{q}_0 \omega_2 \sin(\omega_2 T)\right] \sin(\omega_2 T)
\end{equation}
when written down in terms of
\begin{equation}
    \tilde{q}_0 = q_0 - \frac{F_2}{m \omega_2^2} = \left(\frac{F_2}{m \omega_2^2} - \frac{F_1}{m \omega_1^2}\right) [\cos(\omega_1 T) - 1]
\end{equation}
which is a consequence of the fact that we have not moved our coordinate system into the new center of the harmonic potential.
\section{Frequency Extraction for Optical Dipole Trap}
\makeatletter
\setlength{\@fptop}{0pt}
\setlength{\@fpsep}{8pt}
\setlength{\@fpbot}{0pt}
\makeatother
We simulate the MZI inside the crossed optical dipole potential described in the end matter by first determining the ground state using an imaginary time evolution and afterwards propagating the wavefunction with a standard split-step method. The pulses are implemented using the respective unitary operations, Eq.~\eqref{eq:piHalf} and~\eqref{eq:pi}. Inserting the trap frequencies obtained by Taylor expansion into our analytic model reveals the intervals for $T$ in which the resonances of the two traps are separated in the signal. Likewise, our analytic results for the breathing contribution~\eqref{eq:breathingPhaseFinal} and~\eqref{eq:breathingContrastFinal} can be used to show that the small difference between the transversal trap frequencies does not have a significant impact on the signal of the MZI.

\begin{figure}[p]
    \centering
    \includegraphics[width=510.00108pt,height=626.24928pt]{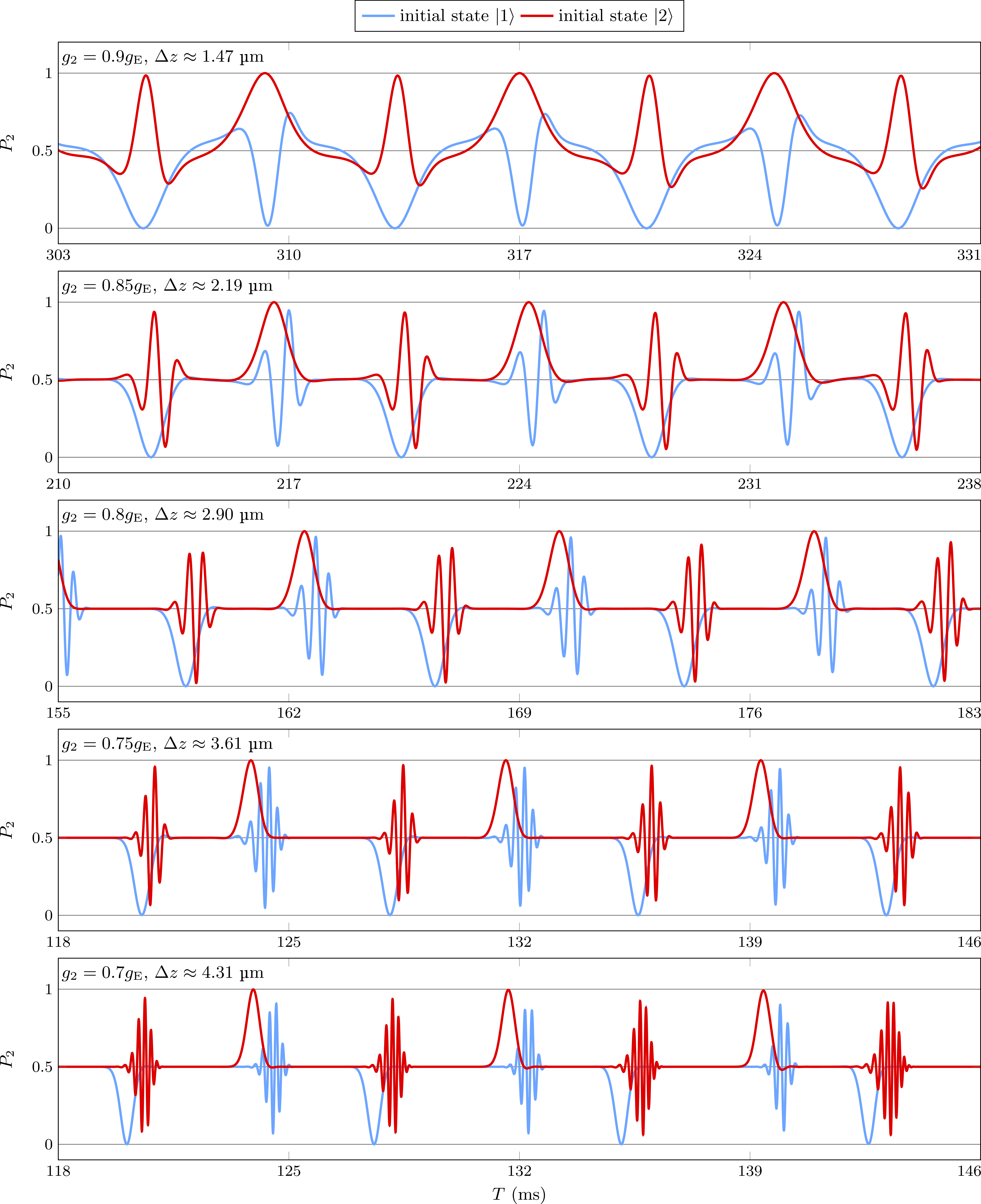}
    \caption{\label{fig:dipoleSignals}
        Interferometer signals $P_2$ as a function of the time between the pulses $T$ obtained by simulating the MZI in a crossed optical dipole trap as discussed in the main text ($\Delta \Phi_\text{P} = 0$).
    }
\end{figure}

The results of the simulations are shown in Fig.~\ref{fig:dipoleSignals}.
For each trap configuration we have simulated the MZI either starting in state $\ket{1}$ (blue) or in state $\ket{2}$ (red). Since we are always recording the population of state $\ket{2}$, the former case results in minima instead of maxima. The figure shows that the resonances become narrower as the separation between the trap centers increases. Similarly, the slope~\eqref{eq:slope} of the phase increases which leads to more rapid oscillations.

Without the influence of transversal breathing, we would expect the oscillating resonances to peak at $P_2(T) = 1$ (or $P_2(T) = 0$), c.f. Fig.~\ref{fig:signal1D}. Then we could simulate the MZI again but with a pulse phase contribution of $\Delta \Phi_\text{P} = \pi / 2$ and use the extraction scheme discussed in the main text. However, the anharmonicities of the optical dipole potential lead to new contributions to the phase of the interferometer $\varphi(T)$. Consequently, we face a similar problem as with transversal breathing, which is that $\varphi(T) \neq 0$ when $T$ equals the period of the motion.

\begin{figure}[htb]
    \centering
    \includegraphics[width=510.00108pt,height=128.05981pt]{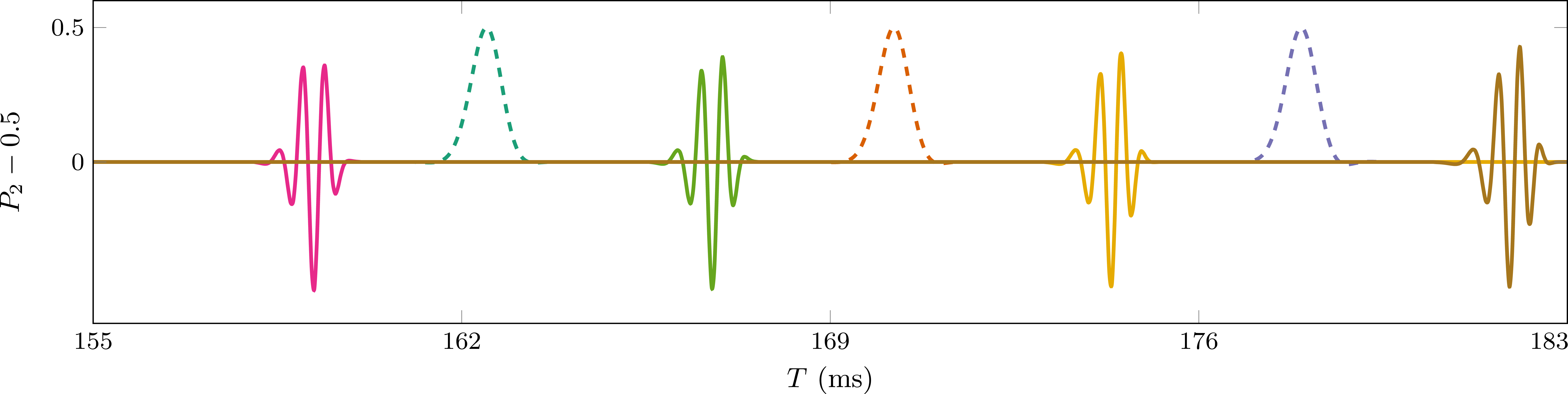}
    \caption{\label{fig:dipoleSeparated}
        Shifted interferometer signal $P_2 - 0.5$ as a function of the time between the pulses $T$ obtained by simulating the MZI in a crossed optical dipole starting in state $\ket{2}$ with $g_2 = 0.8 g_\text{E}$ and $\Delta \Phi_\text{P} = 0$.
    }
\end{figure}

\begin{figure}[htb]
    \centering
    \includegraphics[width=509.9983pt,height=137.34326pt]{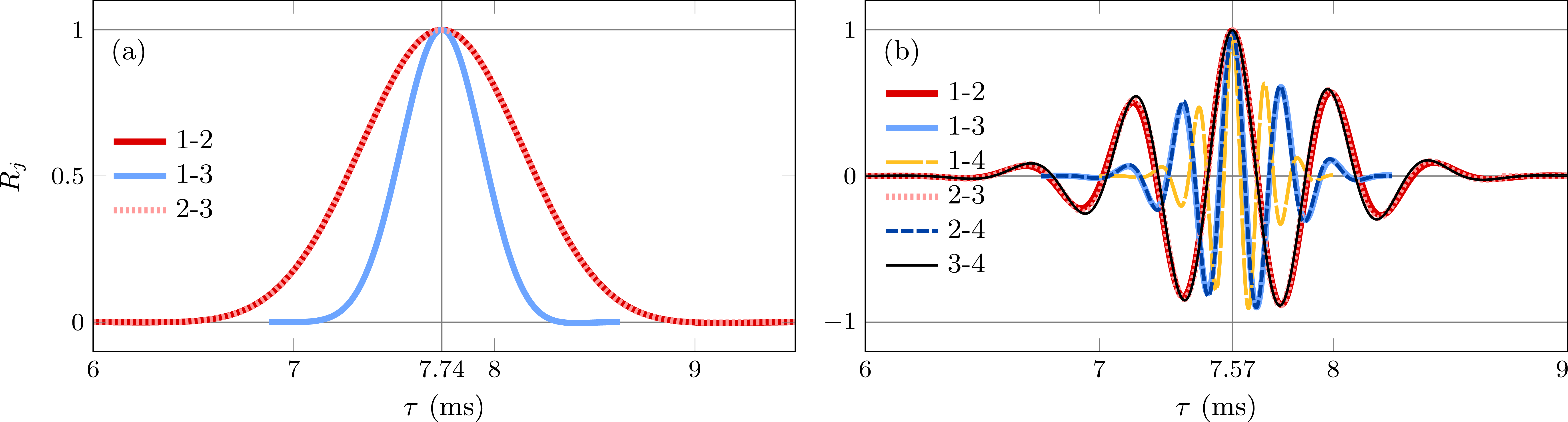}
    \caption{\label{fig:dipoleCorrelations}
        Cross-correlations $R_j$ as a function of the delay $\tau$ for all combinations of the quasi-Gaussian (a) or oscillating (b) resonances in Fig.~\ref{fig:dipoleSeparated}. The legend states the combination of resonances counting from left to right.
    }
\end{figure}

Since the signal is essentially a combination of two periodic processes, other methods to extract the periods can be used. Here we apply the second method mentioned in the main text, which is determining the shift that maximizes the overlap between resonances of the same type. As preparation we split the signal into the individual resonances and shift them by $0.5$ as exemplified in Fig.~\ref{fig:dipoleSeparated} for the MZI starting in state $\ket{2}$ and $g_2 = 0.8 g_\text{E}$.
For each type of resonance (dashed quasi-Gaussian or solid oscillating) a normalized cross-correlation
\begin{equation}
    R_j(\tau) = \frac{\int \mathrm{d}T \ f(T) g(T + j \tau)}{\sqrt{\left[\int \mathrm{d}T \ f^2(T)\right] \left[\int \mathrm{d}T \ g^2(T)\right]}}
\end{equation}
is calculated between all possible combinations of individual resonances. The integer $j$ is set to the number of periods which the resonances are apart. Figure~\ref{fig:dipoleCorrelations}(a) shows the result of calculating the three possible cross-correlations for the quasi-Gaussian resonances in Fig.~\ref{fig:dipoleSeparated}. 
Note that the correlations between the first and third resonance is more narrow because we set $j = 2$. 

When starting the interferometer in state $\ket{2}$, the quasi-Gaussian resonances determine the period of the motion in the potential affecting state $\ket{1}$. Hence, from the average delay $\tau \approx 7.73808~\text{ms}$ at which the correlations in Fig.~\ref{fig:dipoleCorrelations}(a) peak, we obtain $\omega_{1,z} \approx 2 \pi \times 129.231~\text{Hz}$. The process is repeated for the oscillating resonances which yields the correlations in Fig.~\ref{fig:dipoleCorrelations}(b), a average peak delay of $\tau \approx 7.56888~\text{ms}$, and frequency $\omega_{2,z} \approx 2 \pi \times 132.12~\text{Hz}$.
\end{document}